\documentclass[number,sort&compress,5p]{elsarticle}

\begin{document}

\begin{frontmatter}
  \title{Status and recent results of the South Pole Acoustic Test Setup}

  \author[ttk]{Timo Karg}
  \ead{karg@uni-wuppertal.de}
  \author[i3]{for the IceCube collaboration}
 \address[ttk]{Bergische Universit\"at Wuppertal, Fachbereich C ---
    Mathematik und Naturwissenschaften, 42097 Wuppertal, Germany}
  \address[i3]{http://icecube.wisc.edu/}

  \begin{abstract}
    The South Pole Acoustic Test Setup (SPATS) has been deployed to
    study the feasibility of acoustic neutrino detection in Antarctic
    ice around the South Pole. An array of four strings equipped with
    acoustic receivers and transmitters, permanently installed in the
    upper 500~m of boreholes drilled for the IceCube neutrino
    observatory, and a retrievable transmitter that can be used in the
    water filled holes before the installation of the IceCube optical
    strings are used to measure the ice acoustic properties. These
    include the sound speed and its depth dependence, the attenuation
    length, the noise level, and the rate and nature of transient
    background sources in the relevant frequency range from 10~kHz to
    100~kHz. SPATS is operating successfully since January 2007 and
    has been able to either measure or constrain all parameters. We
    present the latest results of SPATS and discuss their implications
    for future acoustic neutrino detection activities in Antarctica.
 \end{abstract}

\end{frontmatter}

\section{Introduction}

The origin and acceleration of cosmic rays to energies beyond
$10^{20}$~eV is one of the big open questions in astroparticle physics
today. Carrying electric charge, cosmic rays are deflected in magnetic
fields during their propagation and possibly do not point back to
their source. However, the cosmic rays being hadrons, interactions in
matter or photon fields at the source should produce high energy
neutrinos that would arrive at earth undeflected and their energy
spectrum will carry valuable information of the physics processes at
their source. For a recent review see
e.g.~Ref.~\cite{Anchordoqui:2010fk}. If the cosmic rays at the highest
energies observed are protons, there additionally is the ``guaranteed
flux'' of cosmogenic neutrinos, produced in the interaction of cosmic
rays with energies above approx.~$10^{19}$~eV with the cosmic
microwave background radiation \cite{Berezinsky:1969fu,
  Engel:2001ff}. Measuring the energy spectrum of cosmogenic neutrinos
will allow one to study different models of cosmological source
evolution. Neutrinos at ultra high energies are also a valuable tool
to study the neutrino-nucleon cross section at high center of mass
energies.

The detection of the small neutrino flux predicted at the highest
energies ($E > 10^{17}$~eV) requires detector target masses of the
order of 100 gigatons, corresponding to 100~km$^3$ of water or
ice. The optical Cherenkov neutrino detection technique, currently
employed in experiments like IceCube, ANTARES, or Baikal, is not
easily scalable from 1~km$^3$-scale telescopes to such large
volumes. Promising techniques, with longer signal attenuation lengths,
allowing for the sparse instrumentation of large natural targets like
Antarctic ice, are the radio and acoustic detection methods. The radio
approach utilizes the Askaryan effect, the coherent emission of radio
waves from the charge asymmetry developing in an electromagnetic
cascade in a dense medium \cite{Askaryan:1962kx}. Acoustic detection
is based on the thermo-acoustic sound emission from a particle cascade
depositing its energy in a very localized volume causing sudden
expansion that propagates as a shock wave perpendicular to the cascade
\cite{Askaryan:1979vn}.

Experiments utilizing the radio technique, the RICE experiment at
South Pole \cite{Kravchenko} being co-deployed with IceCubes
predecessor AMANDA, and the balloon borne ANITA experiment
\cite{Gorham:2010fk} currently deliver the most stringent upper limits
on the cosmogenic neutrino flux.

The South Pole Acoustic Test Setup has been designed to study the
feasibility of acoustic neutrino detection at the South Pole by
measuring the acoustic properties of the ice in the frequency range of
interest from 10~kHz to 100~kHz. These parameters include:

\begin{itemize}
\item The speed of sound and its variation with depth. A gradient in
  the speed of sound depth profile leads to refraction during the
  propagation of acoustic signals over large distances and makes the
  reconstruction of the position of the source more difficult and can
  lead to multiple solutions.
\item The attenuation length and its frequency dependence. This will
  be one of the drivers for the geometry of a future acoustic neutrino
  telescope, determining, together with the noise level, the required
  spacing between sensors for a given lower energy
  threshold. Frequency dependent attenuation will lead to a variation
  of the expected acoustic signal shape with distance.
\item The absolute noise level. Residual acoustic noise in the ice
  will be the other main quantity to determine the energy
  threshold. Its stability with time is important to be able to
  operate a detector with fixed trigger settings.
\item Characterization of any sources of impulsive noise, which might
  be misinterpreted as neutrino signals. To enhance the signal to
  noise ratio it is essential to have a good description of the
  temporal and spatial distribution of transient signals in the ice.
\end{itemize}

Measurement of all these parameters will allow us to get a realistic
sensitivity estimate for a possible future acoustic neutrino telescope
in Antarctic ice.

\section{Detector setup and data taking}

To measure the acoustic properties of the ice at the geographic South
Pole a system of four instrumented vertical lines, called
\textit{strings} A, B, C, and D, have been installed in boreholes of
the IceCube neutrino telescope after deployment of the IceCube optical
modules. Each string holds seven \textit{stages}, a combination of an
acoustic \textit{sensor} and \textit{transmitter} separated by a
distance of approx.~1~m. The horizontal distances between strings
range from 125~m (which is the spacing between IceCube strings) to
543~m. Vertically the depth range from 80~m to 500~m has been
instrumented with increasing spacing of sensors in the deeper ice to
be able to sample the transition from the firn\footnote{The transition
  region from a more loose snow/air mixture at the surface to solid
  ice is called firn. It has a width of 150~m to 200~m.}  region to
the bulk ice. Figure~1 schematically shows the configuration of SPATS
and the depth distribution of the acoustic stages. Strings A to C have
been deployed in January 2007, string D with improved sensors and
transmitters and reaching deeper to 500~m has been installed one year
later in December 2007.

\begin{figure*}[th]
  \centering 
 \includegraphics[width=0.6\textwidth]{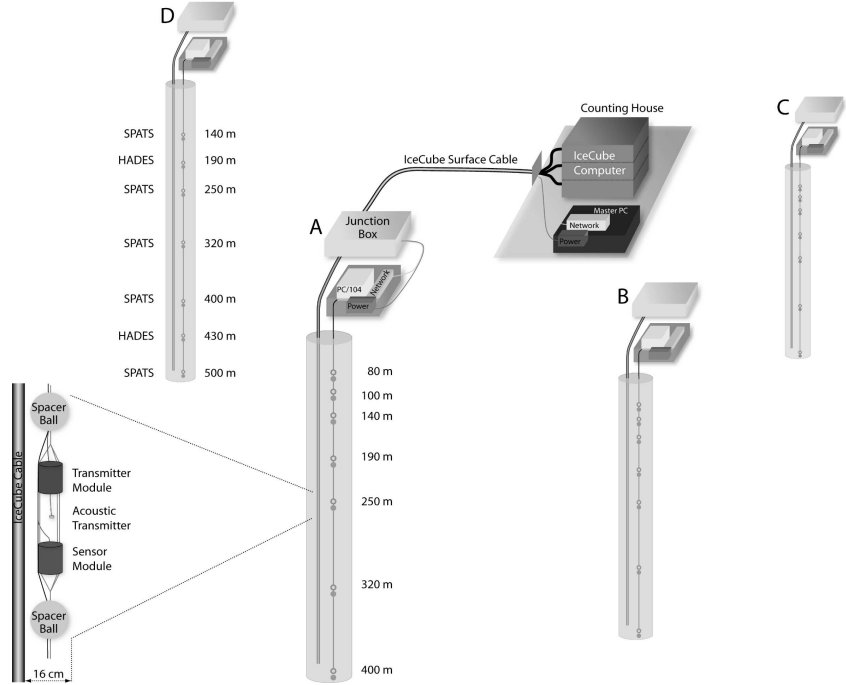}
  \caption{Schematic of SPATS detector setup deployed in IceCube
    boreholes at the South Pole.}
  \label{fig:spats-setup}
\end{figure*}

The SPATS sensors consist of a cylindrical stainless steel pressure
housing with a diameter of 10~cm in which three piezoelectric elements
are mounted to the wall, separated by 120 degrees to get full
azimuthal coverage. A three-stage low noise pre-amplifier with a gain
of $10^4$ is attached directly to the piezoelectric element and the
analogue signal is transmitted to the surface on a shielded twisted
pair cable. The transmitters are comprised of a high voltage pulse
generator protected in a steel housing that drives a ring shaped
piezoelectric element casted in epoxy for protection and mounted below
the housing. The high voltage generator produces pulses with an
amplitude of up to $1.5$~kV and a length of 17~$\mu$s. For string D
the experience gained during the first year of SPATS operation has
been used to improve sensors and transmitters. In the string D sensors
a central metal post that was used to press the piezoelectric elements
to the housing wall was removed and all piezoelectric elements are now
mounted to the wall individually. This has reduced mechanical coupling
and sound transfer between the three individual channels. Further, on
string D at a depth of 190~m and 430~m an alternative type of sensor,
HADES (Hydrophone for Acoustic Detection at South Pole), has been
deployed. For the HADES sensors the piezoelectric element and
pre-amplifier have been moved out of the steel housing and have been
cast in resin and mounted below the housing, similar to the
transmitters. This allows us to study acoustic signals with different
systematics introduced by the sensor. The string D transmitters have
an optimized high voltage circuit design which allows for longer
duration pulses with higher amplitudes.

On the surface, buried in a read-out box in the snow above each
string, an industrial PC, called string-PC, is placed that is used for
digitization, time stamping, and storage of the data. Each string-PC
is connected by a symmetric DSL connection to the SPATS master-PC that
is housed in the IceCube laboratory at South Pole station. The
master-PC collects the data from all four string-PCs, distributes a
GPS timing signal to them, and prepares the data for transfer to the
northern hemisphere via satellite or tape storage. SPATS is operated
in different data taking modes. Of each hour, 45 minutes are used for
transient data taking with three channels of each string, using a
simple single channel threshold trigger. On each trigger signal 5~ms
of data are recorded with a sampling rate of 200~kHz centered around
the triggering sample. The remaining 15 minutes are used for recording
environmental and system health monitoring data. This includes
monitoring of the acoustic noise floor for which each sensor channel
records 100~ms of untriggered data at 200~kHz sampling rate. SPATS is
described in detail in Ref.~\cite{Abdou}.

To increase flexibility in data taking as well as the range of
available transmitter-receiver distances a retrievable transmitter,
called \textit{pinger} has been developed. While constantly emitting
acoustic pulses, it is lowered into the water filled IceCube
boreholes, stopped for five minutes at the SPATS instrumented depths,
and removed before installation of the IceCube optical modules. The
pinger has been deployed in six holes in 2007/2008 where it was
operated with broad band (approx.~15~kHz to 30~kHz) pulses repeated at
1~Hz. The obtained data led to the measurement of the speed of sound
depth profile. In the following IceCube drilling season it was used in
four holes with major technical improvements. Mechanical centralizers
were installed to prevent the pinger from swinging in the hole and
thus stabilize the waveforms. The emitted pulse type was kept the
same, but with an increased repetition rate of 10~Hz. These
modifications, as well as the more favorable positions of the IceCube
holes being drilled with respect to the SPATS array, led to the
determination of the attenuation length. In 2009/2010 the pinger was
modified to emit lower bandwidth pulses at three well defined
frequencies (30, 45, and 60~kHz) and deployed in three boreholes, also
going to deeper depths, down to 1000~m. The measured data is used to
study the frequency dependence of the attenuation length, as well as
the speed of sound on inclined paths.

\section{Results}

\subsection{Speed of sound}

We determined the speed of sound on horizontal paths of 125~m length
for all SPATS instrumented depths from 80~m down to 500~m from the
2007/2008 pinger data. The pinger emission is time-synchronized to a
GPS clock which allows for a precise measurement of the propagation
time from the pinger to the SPATS sensors. In addition to the sound
speed profile for longitudinal acoustic waves ($p$-waves), also a
profile for transverse waves ($s$-waves) could be measured. While
$p$-waves are produced directly by the pinger in the water filled
IceCube borehole, the $s$-waves, which cannot propagate in water, are
generated at the water-ice interface from $p$-waves with non-normal
angle of incidence.

In the top 200~m of the ice shield, the firn region, a strong sound
speed gradient is observed, that is consistent with previous
data. Below the firn our measurement is the first one for $p$-waves in
this depth region, and the first in-situ measurement for
$s$-waves. The resulting sound speed gradient $g$ is compatible with
zero, implying that sound waves propagate without refraction below a
depth of 200~m \cite{Abbasi:2010kx}:

\begin{eqnarray*}
  v_p (375~\mathrm{m}) & = & (3878 \pm 12)~\mathrm{m} \,
  \mathrm{s}^{-1} \\
  g_p & = & (0.087 \pm 0.13)~\mathrm{m} \,
  \mathrm{s}^{-1}~\mathrm{m}^{-1} \\
  v_s (375~\mathrm{m}) & = & (1975.8 \pm 8.0)~\mathrm{m} \,
  \mathrm{s}^{-1} \\
  g_s & = & (0.067 \pm 0.086)~\mathrm{m} \,
  \mathrm{s}^{-1}~\mathrm{m}^{-1}
\end{eqnarray*}

\noindent The largest uncertainty in the measurement results from the
determination of the horizontal distance between emitter and receiver.

The data obtained in the 2009/2010 season with the pinger being
deployed to a depth of 1000~m will enable us to additionally measure
the speed of sound on inclined paths. This will allow us to probe the
fabric of the ice: if there is a preferred crystal orientation in
South Pole ice the measured speed of sound will vary depending on the
direction. For single crystal ice the sound speeds along and
perpendicular to the crystals $c$ axis differ by 4\%
\cite{Petrenko:1999uq} which can be resolved with the precision
achievable with SPATS.

\subsection{Attenuation length}

Measuring the attenuation length requires the comparison of signal
amplitudes or energies after different propagation lengths through the
ice. To achieve this, the 2008/2009 pinger was optimized by mechanical
centralizers to prevent it from swinging during deployment, reducing
pulse to pulse variations caused by different signal transmission
characteristics at the water-ice interface, by using IceCube holes
that were aligned with respect to the SPATS array, thus reducing the
systematic uncertainties due to variations in the azimuthal response
function of our receivers, and by higher emission rates (10~Hz instead
of 1~Hz) to improve the signal to noise ratio.

We analyzed data sets using different sound sources -- the pinger, the
frozen-in SPATS transmitters, and transient signals from freezing
IceCube holes -- to determine the attenuation length. To minimize the
uncertainties due to different sensitivity of the sensors and unknown
angular response, the attenuation length was determined with each
sensor individually placing the sound source at different distances to
the receiver, while trying to keep the same direction seen from the
sensor. All methods consistently deliver an attenuation length of
about 300~m with a 20\% uncertainty \cite{Abbasi:2010fk}. The error
represents the spread between attenuation lengths measured with each
sensor. This result is in strong contradiction with the
phenomenological model prediction of $(9 \pm 3)$~km with absorption as
the dominant mechanism and negligible scattering on ice grains
\cite{Price:2006fk}.

To investigate this discrepancy the pinger has been modified for the
2009/2010 Pole season to allow us to measure the attenuation length at
different frequencies $f$ from 30~kHz to 60~kHz. The result will help
to discriminate between different attenuation mechanisms contributing:
the scattering coefficient is expected to increase with $f^4$ while
the absorption coefficient should be nearly frequency independent. The
modified pinger has successfully been deployed in three IceCube holes
aligned with respect to the SPATS array at horizontal distances
between 180~m and 820~m and delivered high quality data which is
currently being analyzed.

\begin{figure}[ht]
  \centering
  \includegraphics[width=0.45\textwidth]{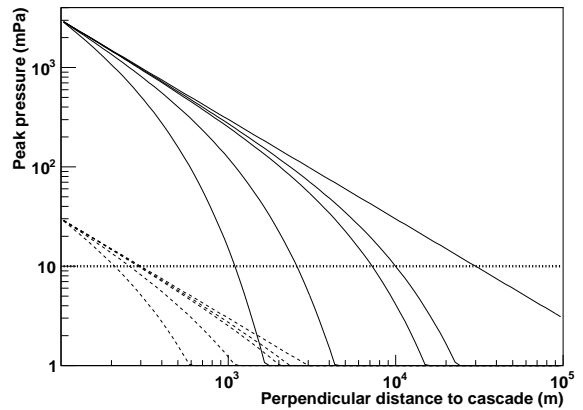}
  \caption{Neutrino signal peak pressure at different perpendicular
    distances to the cascade using a simple model. The solid lines
    correspond to a cascade energy of $10^{20}$~eV, the dashed lines
    correspond to a cascade energy of $10^{18}$~eV. The five lines in
    each set illustrate different attenuation lengths (top to bottom:
    infinity, 9~km, 5~km, 1~km, 300~m). A hypothetical trigger
    threshold of 10~mPa is indicated. For low energies the loss in
    ``visibility range'' due to an attenuation length of 300~m is only
    about a factor of $1.5$ compared to no attenuation.}
  \label{fig:attenuation}
\end{figure}

To illustrate the impact of the shorter-than-expected attenuation
length on acoustic neutrino detection at South Pole let us assume a
simple model for the peak pressure $p_\perp (d)$ in the distance $d$
perpendicular to the center of the axis of a cascade of energy
$E_\mathrm{casc}$. This model approximately reproduces the more
detailed calculations presented for example in \cite{Bevan:2009kx}:

\begin{displaymath}
  p_\perp (d) = 30~\mathrm{mPa} \cdot
  \frac{E_\mathrm{casc}}{10^{18}~\mathrm{eV}}
  \frac{100~\mathrm{m}}{d} \, e^{- (d - 100 \, \mathrm{m}) / \lambda}
\end{displaymath}

\noindent where $\lambda$ is the attenuation
length. Figure~\ref{fig:attenuation} shows $p_\perp (d)$ for two
different cascade energies and different attenuation lengths together
with a trigger threshold of a future hypothetical neutrino telescope
of 10~mPa. It can be seen that for an energy of $10^{18}$~eV, close to
the lower energy threshold where the largest part of the neutrino flux
is expected for e.g.~cosmogenic neutrinos, the pulse amplitude
decrease is dominated by the $1/d$ geometric term and the loss in
``visible range'' is only a factor of $1.5$ from no attenuation to
300~m attenuation length. Only for significantly higher energies
attenuation becomes the dominant loss mechanism, but at these high
energies neutrino signals still can be detected over very large
distances.

\subsection{Absolute noise level}

The high stability of the noise level over time \cite{Karg:2009nx}
makes Antarctic ice a favorable medium compared to (sea) water.
However, the measurement of the absolute noise level, i.e. the
acoustic pressure incident on the sensor, turns out to be one the most
persistent problems within the SPATS project.

The electric noise signal measured at the input of the ADC is a
superposition of electronic self noise of the sensor, electromagnetic
interference picked up during the analogue data transmission from the
sensor to the surface over a twisted pair connection, and possible
acoustic noise contributions in the ice. While all SPATS sensors have
been calibrated in the lab in water at $0^\circ$C before deployment,
it is not at all clear that their sensitivity will remain the same in
South Pole ice at temperatures of $-50^\circ$C, increased static
pressure, and the different acoustic impedance matching at the
ice-sensor, as compared to the water-sensor, interface. While the lack
of a standardized sound source for ice makes in situ calibration of
the sensors impossible, the influence of the different environmental
effects on the sensitivity can be studied separately in the lab
\cite{Karg:2009kx}. The sensitivity of an exemplary SPATS sensor in
air increased by a factor of $1.5 \pm 0.2$ when cooled down to
$-50^\circ$C. For the same sensor, no systematic change with ambient
pressure (in water at room temperature) was observed in the range up
to 100~bar. The sensitivity variation due to pressure changes was
determined to be less than 30\%. The variation of sensor response due
to the coupling to a different medium is currently under study in a
$3~\mathrm{m}^3$ block of clear ice produced in a tank of the IceTop
air shower array placed in a freezing container.

\begin{figure}[ht]
  \centering
  \includegraphics[width=0.45\textwidth]{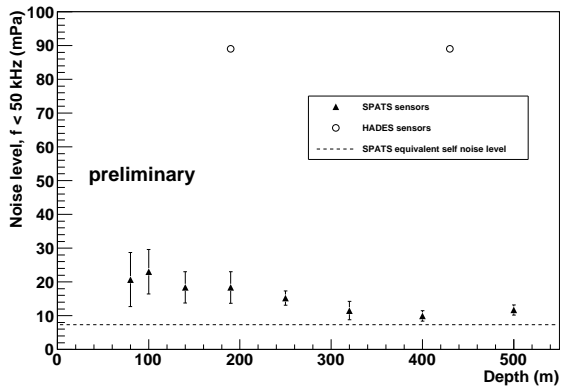}
  \caption{Estimated acoustic noise level at the SPATS instrumented
    depths (see text for details) and mean equivalent self noise
    measured in the lab prior to deployment. The error bars represent
    the spread between different sensor channels. The two HADES
    sensors are dominated by electronic self noise.}
  \label{fig:noise-level}
\end{figure}

Due to the unknown correlation between the different environmental
effects when exposing the sensor to a combination of all of them in
the deep Antarctic ice a measurement of the absolute noise level is
currently not possible. We estimate the noise level at South Pole
under the assumption that the influence of temperature and static
pressure are uncorrelated and thus the in-situ sensitivity is
increased by a factor of $1.5$ compared to the initial laboratory
calibration. To minimize the influence of a possibly modified
frequency response due to the different acoustic coupling we will not
use the sensor's full frequency dependent response, but will use a
mean sensitivity of $2.8$~V/Pa before any corrections, averaged over
all sensor channels and over the relevant frequency range $f <
50$~kHz. Integrating the noise power spectrum of each sensor from 0 to
50~kHz and dividing by the mean sensitivity times the factor of $1.5$
correcting for the temperature yields a noise level for each
channel. Figure~\ref{fig:noise-level} shows the resulting noise level
as function of sensor depth; the error bars indicate the spread
between the up to 12 sensor channels deployed at each depth. Also
indicated is the 7~mPa mean equivalent self noise measured in the lab
prior to deployment. Quadratic subtraction of the equivalent self
noise yields a residual noise that is a superposition of acoustic
noise, electronic noise, and electromagnetic interference. We also
show the values derived for the two HADES sensors which are
unfortunately dominated by electronic self noise.

To obtain a more reliable measurement of the absolute noise level,
including sensitivity variations induced by the coupling of the sensor
to the ice, we plan to deploy two new sensors during the 2010/2011
austral summer. These sensors are equipped with an acoustic
transmitter built into the steel housing, exciting the sensor in a
well defined manner independent of the medium outside. Comparing
signals generated with this transmitter with the sensor being in air,
water, or ice will allow us to follow the evolution of response from
deployment until equilibrium with the bulk ice is reached.

\subsection{Transient noise sources}

\begin{figure}[ht]
  \centering
  \includegraphics[width=0.45\textwidth]{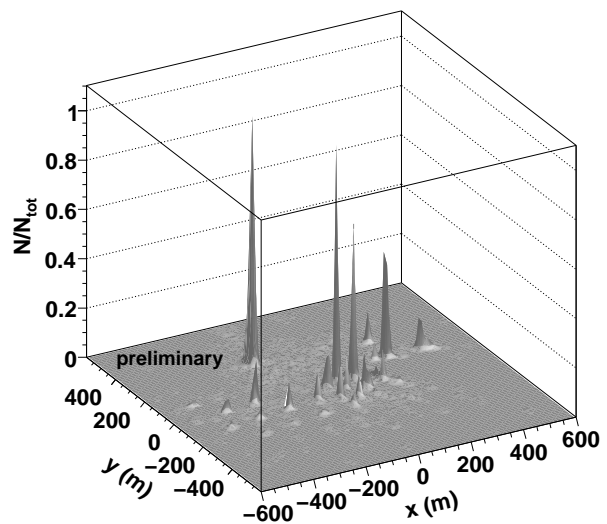}
  \caption{Normalized horizontal distribution of all transient event
    vertices recorded since August 2008. The vertices are well
    localized on the horizontal plane. All observed events originate
    from either the Rodriguez Wells, large caverns melted in the ice
    for water storage during IceCube drilling, or are emitted from the
    re-freezing IceCube holes.}
  \label{fig:transients}
\end{figure}

In transient data taking mode three channels on each SPATS string
record triggered data. Above-threshold events in each single channel
are written to disk with a typical trigger rate of $0.1$~Hz per
channel, a data set that is dominated by single bin excesses from the
tail of the Gaussian noise distribution. A search for coincidences in
time is performed offline with a coincidence window of 200~ms
corresponding to the longest distance across the SPATS array of
approx.~775~m. The vertex of transient events producing triggers on
all four strings is reconstructed using an idealized global
positioning system algorithm. The horizontal positions of all
reconstructed vertices are shown in Fig.~\ref{fig:transients}.

All sources of transient noise are well localized in space and have
been identified as being man made:

\begin{itemize}
\item IceCube boreholes re-freezing after the deployment of the
  optical module produce cracking noise for a period of about 20 days
  after which they quiet down again.
\item Rodriguez Wells are caverns melted into the ice at a depth of
  50~m to 100~m as a water storage reservoir during IceCube
  drilling. With the relocation of the drill camp each austral summer
  a new Rodriguez Well is created and the old wells, with an unknown
  amount of residual water inside, are abandoned to re-freeze.
\end{itemize}

The absence of any transient events observed from locations other than
known sources allows us to set a limit on the flux of ultra high
energy ($E_\nu > 10^{20}$~eV) neutrinos.  Details on transient noise
in SPATS and the neutrino flux limit are presented at this conference
in Ref.~\cite{Berdermann}.

To be able to operate SPATS with lower trigger thresholds -- the
current threshold is set to limit the data volume to the allocated
bandwidth for satellite data transfer from South Pole -- a new data
acquisition software framework has been developed and is currently
being tested in SPATS. This will allow us to have online communication
between the string-PCs and the master-PC and search online for
coincidences of acoustic signals in time, reducing the data volume
being stored.

\section{Future plans}

The ice acoustic properties measured by SPATS allow us to study the
feasibility of acoustic neutrino detection in Antarctic ice. If our
estimations of the absolute noise level are verified by the planned
measurement, the unexpectedly short attenuation length can be
compensated for by glaciophones with low self noise measuring signals
at the limit of acoustic background noise.

Due to the low expected neutrino flux at ultra high energies and the
absence of a natural calibration source, like atmospheric muons and
neutrinos are for optical Cherenkov neutrino telescopes, good
background rejection will be an important topic for a future UHE
neutrino detector. One possibility to achieve is hybrid detection with
at least two different, independent detection systems, reducing
systematic uncertainties and allowing for cross calibration. Two
techniques aiming at large, sparsely instrumented detectors are radio
and acoustic neutrino detection. Both approaches currently favor
instrumentation with a typical horizontal spacing of the order of one
kilometer, which suggests common deployments. While for the radio
Cherenkov cone signature shallow deployments of antennas seem
sufficient, for the acoustic pancake-shaped signal pattern deeper
sensors (down to a few hundred meters) improve the sensitivity to
neutrinos significantly.

The radio detection method intrinsically has a lower energy threshold
than the acoustic technique. However, due to the very low acoustic
transient noise background, at neutrino energies of the order of
$10^{17}$~eV, below the threshold for acoustic neutrino detection, the
trigger from the radio detector could be used to read out all acoustic
sensors. A single pulse in one of the acoustic sensors in temporal
coincidence with the radio signal already can add valuable information
to the significance of the detection. At energies above
approx.~$10^{18}$~eV a standalone acoustic multiplicity trigger may be
feasible.

\begin{figure}[ht]
  \centering
  \includegraphics[width=0.45\textwidth]{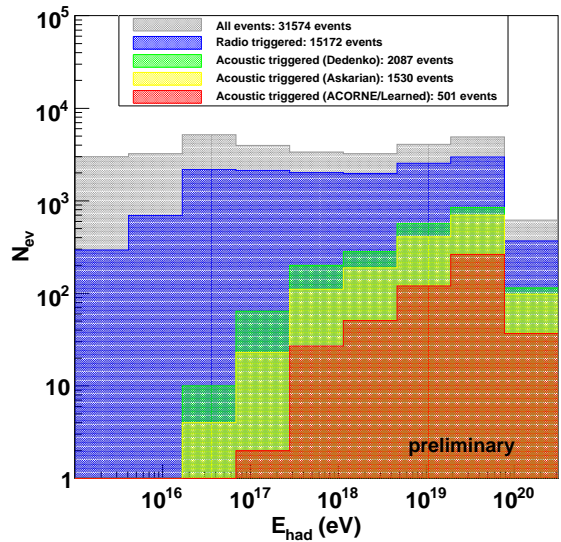}
  \caption{Number of triggered events in a 25~km$^2$ radio-acoustic
    array (see text for details) detected by radio only and producing
    a detectable signal in at least one acoustic sensor for three
    different models of sound emission (from
    \cite{Berdermann:2009uq}).}
 \label{fig:radio-acoustic}
\end{figure}

Simulations of different hybrid radio-acoustic detector arrays
\cite{Berdermann:2009uq} show that for an array of strings deployed on
a quadratic grid with 500~m spacing covering an area of 25~km$^2$ and
deploying several radio antennas and acoustic sensors to a depth of
300~m, between three and ten percent (depending on the acoustic
emission model used) of all neutrino events triggering the radio
detector will also produce detectable signal in at least one acoustic
sensor. Figure~\ref{fig:radio-acoustic} shows the number of simulated
events per energy bin triggering the radio detector and generating a
detectable signal in at least one acoustic sensor for different sound
emission models. Such an array would lead to the hybrid detection of
the order of one cosmogenic neutrino per year, assuming the standard
ESS neutrino flux \cite{Engel:2001ff}.

Two large in-ice radio projects are currently pursued in Antarctica:
the Askaryan Radio Array (ARA) \cite{Hoffman} at South Pole and
ARIANNA \cite{Barwick:2007fk} on the Ross ice shelf. We will study the
feasibility of possible acoustic additions to these projects or future
extensions to them.

\section{Conclusions}

The South Pole Acoustic Test Setup has been operated very successfully
for almost four years now and has nearly completed its task to measure
the acoustic properties of South Pole ice relevant for the acoustic
detection of ultra high energy neutrinos. The sound speed depth
profile has been measured for the first time in the deep ice and for
shear waves and has been found to be consistent with a constant sound
speed below 200~m depth. This allows us to reconstruct the position of
acoustic sources without having refraction effects to be taken into
account which might lead to ambiguous solutions. The attenuation
length has been found to be smaller than expected but the impact of
this result is not considered as a serious problem for future acoustic
neutrino detection activities since the distance over which typical
signals can be detected is only decreased by a factor of about
$1.5$. The exact attenuation mechanism is under study with data from a
dedicated measurement using a retrievable transmitter. The ice at
South Pole is free of transient noise that can accidentally be
misinterpreted as originating from neutrinos. All transient noise
observed originates from well localized, man made sources. The
absolute noise level turns out to be challenging to measure due to the
absence of standardized acoustic sources. Estimates based on
reasonable assumptions yield a noise level comparable to the noise in
calm sea, but being very stable with time. A new measurement is
planned for the 2010/2011 austral summer to considerably reduce the
uncertainties on this last unknown, important quantity.

Combining all results we believe that the acoustic neutrino detection
technique can be a valuable part in a future large neutrino telescope
combining acoustic and radio methods into a hybrid detector.

\section*{Acknowledgements}

We acknowledge the support from the following agencies: U.S.~National
Science Foundation-Office of Polar Programs, U.S.~National Science
Foundation-Physics Division, University of Wisconsin Alumni Research
Foundation, the Grid Laboratory Of Wisconsin (GLOW) grid
infrastructure at the University of Wisconsin -- Madison, the Open
Science Grid (OSG) grid infrastructure; U.S.~Department of Energy, and
National Energy Research Scientific Computing Center, the Louisiana
Optical Network Initiative (LONI) grid computing resources; National
Science and Engineering Research Council of Canada; Swedish Research
Council, Swedish Polar Research Secretariat, Swedish National
Infrastructure for Computing (SNIC), and Knut and Alice Wallenberg
Foundation, Sweden; German Ministry for Education and Research (BMBF),
Deutsche Forschungsgemeinschaft (DFG), Research Department of Plasmas
with Complex Interactions (Bochum), Germany; Fund for Scientific
Research (FNRS-FWO), FWO Odysseus programme, Flanders Institute to
encourage scientific and technological research in industry (IWT),
Belgian Federal Science Policy Office (Belspo); University of Oxford,
United Kingdom; Marsden Fund, New Zealand; Japan Society for Promotion
of Science (JSPS); the Swiss National Science Foundation (SNSF),
Switzerland; A.~Gro{\ss} acknowledges support by the EU Marie Curie
OIF Program; J.~P.~Rodrigues acknowledges support by the Capes
Foundation, Ministry of Education of Brazil.

\section*{References}

\end{document}